\documentstyle[11pt,aaspp4]{article}
\newcommand\etal{{\em et al.~}}

\begin{document}
 
\title{Simultaneous Dual Frequency Observations of Giant Pulses from
the Crab Pulsar}
 
\author{Shauna Sallmen \& D. C. Backer}
\affil{University of California, Berkeley, CA 94720-3411}
\affil{ssallmen@astro.berkeley.edu \& dbacker@astro.berkeley.edu}
\author{T. H. Hankins}
\affil{New Mexico Institute of Mining \& Technology, Socorro, NM 87801; thankins@aoc.nrao.edu}
\author{D. Moffett}
\affil{Physics Department, University of Tasmania, GPO Box 252-21, Hobart, Tasmania 7001, Australia; David.Moffett@utas.edu.au}
\author{S. Lundgren}
\affil{MS4051, Lockheed Martin Astronautics, P.O. Box 179, Denver, CO 80201-0179; Scott.C.Lundgren@lmco.com}

\begin{abstract}

Simultaneous measurements of giant pulses from the Crab pulsar
were taken at two widely spaced frequencies using the real-time
detection of a giant pulse at 1.4\,GHz at the Very Large Array to
trigger the observation of that same pulse 
at 0.6\,GHz at a 25-m telescope in Green Bank, WV.
Interstellar dispersion of the signals provided the necessary time to 
communicate the trigger across the country {\em via} the  Internet.
About 70\% of the pulses are seen at both 1.4\,GHz and 0.6\,GHz,
implying an emission mechanism bandwidth of {\em at least} 0.8\,GHz at 1\,GHz
for pulse structure on time scales of one to ten microseconds.
The giant pulse spectral indices fall between $-2.2$ and $-4.9$, which may
be compared to the average main pulse value for this pulsar of $-3.0$. 

The arrival times at both frequencies display a jitter of 100\,$\mu$s within
the window defined by the average main pulse profile and are tightly 
correlated. This tight correlation places limits on both the emission 
mechanism and on frequency dependent propagation within the magnetosphere.

At 1.4\,GHz the giant pulses are resolved into several, closely spaced
components.  Simultaneous observations at 1.4\,GHz and 4.9\,GHz 
show that the component splitting is frequency independent.
We conclude that the multiplicity of components is intrinsic to the emission
from the pulsar, and reject the hypothesis that this is the result of
multiple imaging as the signal propagates through the 
perturbed thermal plasma in the surrounding nebula. 

At both 1.4\,GHz and 0.6\,GHz the pulses are characterized by a fast 
rise time and an exponential decay time which are correlated. 
At 0.6\,GHz the rise time is not 
resolved. The rise and fall times of the 1.4-GHz components 
vary from component to component and from pulse to pulse. 
The pulse broadening with its exponential decay form 
is most likely the result of multipath propagation in intervening ionized 
gas. These decay times, and that seen in contemporaneous
0.3-GHz average pulse data, are large compared to normal conditions for
the Crab pulsar. The most likely location for the perturbed plasma
is the interface region between the pulsar-driven synchrotron nebula and
the overlying supernova ejecta.

\end{abstract}


\newpage

\section{Introduction}
\label{secintro}

The Crab pulsar was discovered in 1968 by the detection of its
extremely strong individual pulses (\cite{sr68}).
Such pulses, which are 100's of times stronger than the average, 
are not seen in most pulsars.  The properties of these giant pulses
have been explored for many years ({\em e.g.},  \cite{hcr70}; \cite{ss70};
\cite{fb90}; \cite{lcu+95}).  Giant pulses in the Crab pulsar occur at all 
radio frequencies, but only {at the rotational phase of}  the main pulse 
and interpulse components.  These two components have counterpart nonthermal 
emission at high frequencies -- from the infrared to gamma ray energies 
-- and may be associated with the outer voltage gaps in the pulsar 
magnetosphere (\cite{ry95}). 
Giant pulses are not seen in the radio precursor nor at the phases 
of the high radio frequency components recently described by Moffett and 
Hankins (1996)\nocite{mh96}. The radio precursor is identified 
as being more typical of the emission from a conventional pulsar and 
is believed to originate at, and be aligned with, the magnetic pole. 

Lundgren \etal (1995)\nocite{lcu+95} found that two separate distributions were
required to describe the fluctuations of single pulse energies\footnote{
Pulsar emission profiles are generally given in units of {\it flux} (Jy) even
though in the context of rotating neutron stars one actually samples a one dimensional
cut of the {\it specific intensity} pattern (Jy sr$^{-1}$). The integral of
emission over a pulse component in the latter case would be its flux, while
in the former and conventional case one quantifies the integrated component
emission in units of energy (Jy-s).
}
for the Crab pulsar main
pulse and interpulse components at 0.8\,GHz.  About 2.5\% of the pulses lie 
in the giant pulse distribution with a power law slope extending to high 
values and a low energy cutoff of 20 times the average of all pulse energies.
The distinct distributions suggest different emission mechanisms
for the giant and weak pulses
and possibly different emission locations within the magnetosphere.
However, the lack of an offset in the timing residuals between giant pulses 
and the average pulse profile (\cite{lun94}; for opposing 
evidence see \cite{fb90}) suggests that the emission region is the same.

The frequency of occurrence of pulses ($f_\circ$) 
with energy greater than 20 times the average
increases with frequency, from $10^{-4}$ at 0.146\,GHz (\cite{ag72}), 
to 0.025 at 0.8\,GHz (\cite{lcu+95}).  
The contribution of the giant pulses to the average energy of all pulses also
increases with radio frequency, although not as quickly.
The probability distribution of the giant pulse energies can be written as
$$P(E_{\rm GP}>E_{\rm o})=f_{\rm o}\left ({E_{\rm o}\over E_{\rm min}}\right )^{-\alpha},$$ 
where $f_{\rm o}$ is the frequency of occurrence of the giant pulses, and
$E_{\rm min}$ is the minimum energy. Correspondingly the probability density
function is
$$p(E_{\rm GP})={f_{\rm o}\alpha\over E_{\rm min}}\left ({E_{\rm GP}\over 
E_{\rm min}}\right )^{-\alpha-1},$$
and the corresponding mean giant pulse energy averaged over all pulses is 
${f_{\rm o}\alpha\over (\alpha-1)}E_{\rm min}$.  The probability distribution 
$P$ has a slope $\alpha = 2.3\pm 0.15$ at 0.8\,GHz (\cite{lcu+95}),
and $\alpha=2.5$ with significant errors at 0.146\,GHz (\cite{ag72}).  
At 1.4\,GHz and 0.43\,GHz the overall slope is roughly consistent with these, 
but is not the same for all energies (\cite{fb90}; \cite{m97}).
Using the scaling law $\alpha \approx 2.5$ at all radio frequencies below 
0.8\,GHz, we find 
that the contribution of giant pulses with energy more than 20 times the average
of all pulses, $E_{\rm GP}>20\,E_{\rm avg}$, is 89\% of the average energy 
at 0.8\,GHz (\cite{lcu+95}), 9\% at 0.43\,GHz (\cite{fb90}),
and only about 1\% at 0.146\,GHz (\cite{ag72}).
At 1.4\,GHz, a similarly large fraction of the energy comes from the 
approximately 2\% of pulses with greater than 20 times the average energy, 
although a single slope $\alpha$ does not accurately describe the distribution.

There is no evidence of increased flux density in pulses near the giant pulses
(\cite{ssp71}; \cite{lun94}), nor is there any correlation between giant pulses.  
{We note that many pulsars do show pulse to pulse correlation 
indicating a memory process with a duration of many rotational periods.}
The timescale of giant pulses is, in contrast, less than a single period.
In addition, the time 
separation distribution for giant pulses is consistent with Poisson process (\cite{lun94}). 

Despite all these studies, the emission bandwidth of the giant pulses has
been poorly determined.  Comella \etal (1969)\nocite{cc+69} found that 50\%
of giant pulses were seen simultaneously at 0.074\,GHz and 0.111\,GHz.  Goldstein
\& Meisel (1969)\nocite{gm69} also found that some but not all pulses 
were correlated between 0.112\,GHz and 0.170\,GHz.  
Sutton \etal (1971)\nocite{ssp71} noted that there was no
evidence that the largest pulses at 0.16\,GHz and 0.43\,GHz were correlated.  
Heiles \& Rankin (1971)\nocite{hr71} observed giant pulses 
simultaneously at 0.318\,GHz and 0.111\,GHz, for a bandwidth spread of about 3:1.  
They found that pulses classified as giant at one frequency were stronger 
than the average at the other, but not usually classified as giant.  
Much more recently, Moffett (1997)\nocite{m97} reported that fully 90\% 
of the giant pulses detected at 4.9\,GHz were also detected at 1.4\,GHz,
implying an emission bandwidth of 3.5\,GHz at high radio frequencies.
In this paper, we report on giant pulses observed
simultaneously at 1.4\,GHz and 0.6\,GHz to explore the correlation
in this intermediate range of frequencies.  Section \ref{sec:gpobs} 
describes the observations, while analysis of the simultaneous pulses 
lies in Section \ref{sec:gpanal}.

\section{Observations}
\label{sec:gpobs}
The data shown here were  recorded on 1996 May 21 at UT 
$17^{\rm h}45^{\rm m}-19^{\rm h}15^{\rm m}$  (1.4/0.6 GHz)
and 1996 Oct 12 at UT $11^{\rm h}30^{\rm m}-12^{\rm h}05^{\rm m}$
(4.9/1.4 GHz). We also refer to observations at 1.4-GHz earlier
in 1996 and in 1997 November.
The 1.4-GHz and 4.9-GHz data were taken at the NRAO Very Large Array 
(VLA).\footnote{The National Radio Astronomy Observatory is a facility
of the National Science Foundation operated under cooperative agreement 
by Associated Universities, Inc.} 
For the simultaneous 1.4/0.6-GHz observations  all 27 VLA antennas 
were phased to create the equivalent {sensitivity} of 
a 130-m antenna, while the 0.6-GHz data
were taken using a 25-m telescope at the NRAO Green Bank, WV, site.  
For the simultaneous 4.9/1.4-GHz observations the VLA was split 
into two equal subarrays.

The peculiar phases of each antenna at the VLA were determined by observing 
a standard point-source calibrator. These phases were then applied to 
the antennas to synthesize a pencil beam pointed at the Crab pulsar, 
which essentially resolves out the bright Crab Nebula and vastly 
improves the signal to noise ratio compared to a single-dish antenna. 
The received voltages from each antenna are summed to form orthogonally 
circularly polarized 50-MHz bandwidth signals centered at precisely
4.8851\,GHz or 1.4351\,GHz. 
The signals are then detected and summed with a 100-$\mu$s time constant.  
The detector rms noise power was determined using an rms to DC converter.
A detector threshold was set at either 5 or 6 times the running average 
of this rms noise level. Pulses that exceeded this threshold generated
a trigger pulse that was sent to the data recorder, and were then
saved to disk and archived to tape using the full 50-MHz bandwidth. 
In an off-line computer the data were coherently dedispersed using the 
method developed by Hankins (1971) \nocite{han71} and described by Hankins 
and Rickett (1975)\nocite{hr75}. Although the ultimate time resolution of 
the dedispersed data is 10\,ns, for the analyses described here the data 
were smoothed to 0.5--1.0\,$\mu$s after software detection.

The two linearly polarized signals centered at exactly
0.610\,GHz were converted to 90\,MHz and 
110\,MHz center frequencies, respectively. The two intermediate frequency
signals were then summed and transmitted on a single fiber optic link from the 
25-m telescope to the Green Bank--Berkeley Pulsar Processor (GBPP)\footnote{
A partial technical description of the GBPP is given in (\cite{b97}). }
which was located at the 140-ft telescope.
The GBPP converted the signals to baseband, split these into
32 0.5-MHz channels, and dedispersed the pulsar signal in each channel 
by (de-)convolution in the time domain.  The dispersion delay 
across the 16-MHz bandwidth of the GBPP at 0.6\,GHz is about 33\,ms, 
or one pulse period for the dispersion measure of the Crab pulsar 
{(${\rm DM}\approx 56.8$ pc\,cm$^{-3}$)}. Full Stokes information for
982 samples across the pulsar period was recorded with an accurate UTC 
start time for each pulse. The 25-m telescope also monitors the Crab
pulsar daily with 0.327-GHz observations which are valuable for their
sensitivity to scatter broadening by intervening plasma.

The 1.4/0.6-GHz part of this experiment utilized the difference in 
pulse arrival time between the two frequencies due to interstellar 
dispersion to provide the time interval needed to communicate the 
trigger information between the sites.
At the VLA, the same trigger pulse that was sent to the data recorder
was also sent, as an interrupt, to the SUN workstation used for experiment
control and recording.  The program that received the interrupt  
had a socket link open over the Internet
to a slave program running on another SUN workstation 
in Green Bank, and communicated the 1.4356-GHz arrival time of the pulse to 
Green Bank.

We arranged for the GBPP to begin taking data just before the giant pulse 
reached the upper edge of the 0.6-GHz band, in order to obtain data for 
the same main pulse across the entire band. The dispersion time delay 
between 1.4351\,GHz and 0.618\,GHz (the top of our 0.6-GHz band) allowed 
a half second (0.503\,s) to arrange this.  The SUN 
workstations at the two sites were synchronized to the local versions of 
UTC which were derived from accurate atomic clocks using the {\tt xntp} 
protocol. Although both remote sites had 56-kB links to the  Internet, 
the typical Internet transfer time delay was 200\,ms during our observation. 
The program running in Green Bank received the trigger message with its VLA 
time stamp, calculated the  transit time of the trigger, and compared that
to the dispersion delay difference of 0.503\,s.  
In addition to this delay, the program included other factors such as the
difference in pulse arrival due to the separation between observing sites 
on the Earth and the latency in the GBPP hardware, both of which were of order 
1--3\,ms.   
{ If sufficient time remained, the program waited until the appropriate time
and issued a trigger to the GBPP  {\em via} the SUN parallel port
to take data for the next pulse period.}
Owing to the slow rate of data transfer from the GBPP, it could only accept such 
a command approximately every 12 s.  Some {VLA-initiated} triggers were 
therefore missed by the GBPP.

For the 4.9/1.4-GHz measurements the recording system was triggered by a 
4.9-GHz pulse, and then automatically triggered again after the appropriate 
dispersion delay to record the 1.4-GHz signals.

The Stokes parameters for the high-time resolution data from the VLA
were formed from the dedispersed voltages. 
The necessary 90$^\circ$ phase shift was obtained using an finite impulse
response approximation 
to the Hilbert Transform and was applied to the right circular polarization 
signal before forming the Stokes parameters. 
No instrumental polarization corrections were made other than bandpass 
leveling and gain matching. Concurrent calibration (Moffett 1997) 
has shown that the polarization cross-coupling is less than 10\% for
the VLA phase array. The polarization error then is comparable
with the radiometer uncertainty imposed by the limited number of degrees of
freedom in the data ($\sigma_I/I = (4\Delta \nu \,\Delta \tau)^{-1/2} = 0.1)$.

The polarization profiles at 0.6\,GHz were calibrated using factors
derived from pulsed noise observations.  The receiver introduces a
relative phase between the two linear polarizations which couples the 
Stokes parameters $U$ and $V$.
This phase was determined and removed using nearby observations of 
the Vela pulsar, and comparison to a template polarization profile.
No attempt has been
made to remove coupling between the two polarizations. The error in polarization
due to improper calibration is estimated at 10\%. 
For each pulse, the relative dispersion between the 32 channels was 
removed, and the resultant data were summed over channels, after first
removing the effects of the pulsar's rotation measure (${\rm RM}=-42.3$ rad~m$^{-2}$) 
across the band. The unknown Faraday rotation from the ionosphere  
causes negligible rotation of 1 to 10 degrees across the total band. 

\section{Analysis}
\label{sec:gpanal}

\subsection{Wide Bandwidth Correlation}

Of the 85 trigger events initiated at the VLA a total of 77 events
reached Green Bank within the required time, and were accepted by the GBPP. 
The {\tt xntp} protocol requires 24 hours to stabilize to the accuracy
required by our experiment.  {The minimum time stabilization period} 
was not available for the SUN
at the VLA.  Consequently the VLA clock used to identify the time at
which the trigger was sent drifted by a small amount.
We are currently certain that the received trigger events allowed
capture of the correct period in the GBPP for 42 pulses.  

The arrival times and pulse energy amplitudes were determined at both 
frequencies for
each of these pulses.  The 0.6-GHz arrival times were determined by
cross-correlation with a model template, which consisted of a single-sided 
exponential with a decay time scale of 3 time bins, about 100\,$\mu$s.  
Owing to complex structure discussed below,
the 1.4-GHz pulse arrival times were obtained by computing the location of the 
centroid of the pulses. Pulse amplitudes in units of Jy-s
were determined by removing an ``off-pulse'' baseline, and then summing the 
flux in the time bins that comprise the ``on-pulse'' window.  The measurement
uncertainty of these amplitudes was determined from the ``off-pulse'' noise
distribution.

We definitely detected 29 of the 42 {correctly timed} pulses at both radio
frequencies.  These detections correspond to a 0.6-GHz pulse energy threshold 
of about 4.5 times the typical measurement uncertainty, or 0.075 Jy-s 
using 0.14 K Jy$^{-1}$ for the 25-m telescope.  This gain factor was determined 
using on and off measurements of the Crab nebula (which is 1208 Jy at 
0.6\,GHz), and has an estimated {uncertainty} of 50\%.
We conclude that about 70\% of the pulses are detected at both frequencies.
This statistic is used to discuss the spectral index distribution below.  

The data for a single giant pulse at 1.4\,GHz and 0.6\,GHz are
displayed in Figures \ref{vlabest} and \ref{gbbest},
respectively.  This pulse is strongly polarized at both frequencies, 
although fully two thirds of the giant pulses at the lower frequency are 
consistent with zero polarization.
The 0.6-GHz data has a low number of degrees of freedom, and so the
{polarization estimation uncertainty} is about 5\%.   
At 1.4\,GHz, the typical polarization is about 8\% 
although at least one pulse is 50\% polarized. 
The position angle of the linear polarization generally varies significantly
across the pulse, as is seen in Figure \ref{vlabest}.

 
The arrival times for the 29 giant pulses detected at the two frequencies 
were separately compared to a single 
model for this pulsar using the TEMPO program developed for pulsar timing (\cite{tw89}). 
For each radio frequency, the arrival times are well-represented by the
model, leaving timing residuals of order $\pm 100\,\mu$s. The residuals
are comparable
to the pulse width of the average profile during periods of low scattering, 
which is $275 \pm 50\,\mu$s (FWHM) at 0.6\,GHz, and $257 \pm 50\,\mu$s at 
1.4\,GHz.  These widths were estimated using GBPP data obtained with the
25-m and 140-ft telescopes, respectively.

The timing residuals for 1.4\,GHz and 0.6\,GHz are plotted against one another in 
the top panel of Figure \ref{resids}, which shows that they are highly correlated.  
The solid line has a slope of
one and goes through the origin.  In order for the points to fall along 
this line, the 1.049-ms digital latency of the GBPP and the 235.42-$\mu$s
latency of the VLA samplers and delay lines were removed, and a further fit for
dispersion measure was done in TEMPO. The derived DM is
56.830 pc cm$^{-3}$ although systematic errors may remain in the arrival
times from the two sites.  
The bottom panel displays the same 1.4-GHz residuals with the
solid line removed.


Eilek (1996) \nocite{eil96} 
has shown that the dispersion law in the polar cap is 
proportional to $\nu^{-1}$, as opposed to $\nu^{-2}$ for 
the cold interstellar medium (ISM). No systematic trends remain
in the data in the lower panel of Figure \ref{resids}, indicating that 
systematic variations with pulse phase 
are less than $\pm 15\,\mu$s between our two bands.  This places limits
on the differential effects of propagation through the magnetosphere.
Geometrically, emission in the two bands must originate within 0.16$^\circ$ of
rotational phase, or a range of 4.5\,km in altitude.
It would have been possible to have correlated emission from subpulses
at different pulse longitudes at each frequency.  In this case, the radiation
at the two frequencies need not have come from the same radiating unit of
charges.  The observed rms jitter in arrival time at either frequency is 
$\approx$100\,$\mu$s, 
so the fact that the difference between the residuals has such a small 
dispersion indicates that the emission must be from the same radiating unit at
both frequencies.
This means that at least 70\% of the giant pulses must have a bandwidth 
of {\em at least} 0.8\,GHz at 1\,GHz.  The emission is clearly broadband 
for these cases.\footnote{ 
Note that in this paper the term ``broadband'' connotes simultaneous
emission over wide range of radio frequencies
with a ratio of amplitudes comparable to that of the
average pulse. However, the observed emission has a very steep 
spectrum, even if weighted by frequency
to obtain total power, and is therefore narrow band in an absolute sense.}

\subsection{Pulse Shape Model}
\label{secshape}

The giant pulses at 0.6\,GHz all have profiles that display a fast rise 
followed by an exponential decay, similar to the profile shown in Figure
\ref{gbbest}.  The exponential decay time scale is 
$\tau_{\rm ISS}(0.6\,{\rm GHz})= 95 \pm 5 \mu$s.
Monitoring of the exponential broadening of the average pulse profile
at 0.3\,GHz using the 25-m telescope (\cite{bw96}) provides a
contemporaneous decay time scale of $\tau_{\rm ISS}
(0.3\,{\rm GHz}) = 1.3 \pm 0.2$\,ms. 
The 0.3-GHz and 0.6-GHz pulse broadening time scales
are consistent with the $\nu^{-4\,{\rm to}\,-4.4}$ dependence expected from scattering
by an intervening, turbulent plasma screen.
At the time of these observations the Crab was undergoing a period of 
unusually large scattering.  The contemporaneous value of $\tau_{\rm ISS}
(0.3\,{\rm GHz})$ may be compared to 0.28\,ms at an earlier epoch.
Enhanced scattering of the Crab
pulsar radiation also occurred in 1975 
(\cite{ir77}; \cite{lt75}).
A likely site of the perturbed plasma that causes the
enhanced pulse broadening is in the interface between the synchrotron nebula
and the supernova ejecta (\cite{hetal96}). This region is proposed 
because such enhanced scattering events are not seen for any other pulsar. 
Furthermore, this location is known to contain dense thermal plasma 
with structure on a variety of length scales. 
The transverse velocity of the line
of sight with respect to  the interface material is sufficient
to bring occasional regions which produce strong perturbations into
view. The fast rise of the 0.6-GHz giant pulses
indicates that any intrinsic time scale of the pulse 
is unresolved, {\lower.5ex\hbox{$\; \buildrel < \over \sim \;$}}10\,$\mu$s. 

The giant pulses at 1.4\,GHz have a wide variety of shapes.  Figures 
\ref{pulse69} and \ref{pulse56}
display two further single pulse profiles. The first pulse 
is extremely narrow and is dominated by a single 
component, while the second 
has several components contributing to the emission.  The darker solid
line is a fit to the data where the flux model $S(t)$ consists of up to six
components and is of the form 
\begin{equation}
S(t) = \sum_{i=1}^n  a_{1_i} (t-a_{2_i}) e^{-(t-a_{2_i})/a_{3_i}},~~~n\leq 6.
\label{eq:xfred_model}
\end{equation}
These components rise to their peak in a time $a_3$, fall by $e^{-1}$ 
in a further $2.15~a_3$, and have a pulse energy amplitude of $A=a_1 a_3^2$.  
At 1.4\,GHz, the majority of giant pulse components are well represented by 
this model, with widths $a_3$ ranging from 1.2\,$\mu$s to 10\,$\mu$s. 
Components with widely varying decay time scales may be superposed within
a single giant pulse, as shown in Figure \ref{pulse56}.
The narrow component of the 
giant pulse shown in Figure \ref{pulse69} has a rise time of
1.2\,$\mu$s, and a decay time scale of 2.5\,$\mu$s.  The weak and broad
second component in Figure \ref{pulse69}
is clearly necessary to account for the emission on the trailing edge of the 
pulse, which does not follow an exponential tail.  


\subsection{Pulse Shape Interpretation}
\label{secinterp}

The 1.4-GHz observations raise fundamental issues about the nature
of the giant pulse emission. Is the 
multiplicity of the components and the characteristic shape of the components a 
result of propagation through a turbulent screen, or are these effects
intrinsic to pulse formation and propagation in the pulsar magnetosphere,
or is there a mixture of effects present?
Two additional results from further observations at the VLA 
and the 0.3-GHz monitoring in Green Bank
provide important constraints to aid in answering these questions.

First, our broadening measurement of 1.3\,ms at 0.3\,GHz scales 
to $1.9-3.5~\mu$s at 1.4\,GHz using the $\nu^{-4{\rm~to~}-4.4}$ dependence expected 
for a spectrum of turbulence filling the intervening screen. This
range is consistent with the typical {\it minimum} broadening 
time of approximately 2.5\,$\mu$s that we measured for the giant pulses at 1.4\,GHz. 
During a later epoch (1997 Nov 26) when the scattering broadening of the
average pulse at 0.3\,GHz
had increased dramatically by a factor of about 5, the minimum pulse widths 
of the 1.4-GHz giant pulses increased by a similar factor. At an earlier
epoch in 1996 both the 0.3-GHz average pulse broadening and the minimum
width of the 1.4-GHz giant pulses were reduced. 

The second observational constraint on our interpretation
of the 1.4-GHz giant pulse shapes comes from consideration of the 
multiple component spacing.  These do not scale in the same way as the component 
broadening.  During an earlier epoch of low 0.3-GHz scattering, the typical 
1.4-GHz component spacing is similar to that for 1996 May 21.  
In addition, simultaneous VLA measurements of the giant pulses at 1.4\,GHz and
4.9\,GHz (1996 Oct 12) show a frequency {\it independence} of the 
multiple component spacing \nocite{hm98}
(Hankins \& Moffett 1998).  At 4.9\,GHz the pulse components  are intrinsically 
very short, typically 0.1 to 0.4\,$\mu$s wide, while the corresponding
components are broadened
at 1.4\,GHz.  Simultaneous 4.9-GHz and 1.4-GHz measurements of a single 
giant pulse are shown in Figures \ref{5GHz}a and \ref{5GHz}b.  The separation
of the two main components is similar at both frequencies, in that the 
onsets of the pulse components (the fiducial points in the fits discussed 
in section \ref{secshape}) are separated by the same amount.  
A similar conclusion of frequency independence of structure
can be inferred from the 1.4\,GHz and 0.6\,GHz measurements of 1996 May 21;
that is, 1.4-GHz giant pulses with multiple components
have a spread of tens of microseconds, which is consistent with the limit
on structure at 0.6\,GHz.   

The correlation of broadening time scales over the range from 
0.3\,GHz to 1.4\,GHz provides support for diffractive scattering in 
the nebular material.  Although the simple thin screen prediction for
interstellar scattering is a single sided exponential, $\exp(-t)$, models
including two widely separated screens or a single thick screen both predict
pulse shapes similar to the observed $(t/a_3)\exp(-t/a_3)$ form of the 
components (\cite{wil74}). Isaacman \& Rankin (1977) \nocite{ir77} have
derived parameters for a two screen model from studies of
the average pulse profile of the Crab pulsar.

A scatter plot of the 1.4-GHz component energies,  $A=a_1 a_3^2$ from the 
fitted parameters in Equation \ref{eq:xfred_model}, {\em vs.} component width, 
$a_3$, shows that the energies are independent of the pulse width, have an 
average of about $6.3 \times 10^{-3}$\,Jy-s, and are scattered over two orders 
of magnitude. This means that the peak flux is approximately inversely 
proportional to the width; stronger pulses are narrower. The data from other 
observing sessions show that although the broadening times change, 
the pulse energies remain within the same range; {\em i.e.}, 
when the scattering times are longer, the peak amplitudes are 
correspondingly smaller.
This multi-epoch result supports the hypothesis that, in spite of questions 
about the pulse component shape and
the frequency dependence of the exponential decay time scale, the shape
of the components at 1.4\,GHz is most likely the result of scattering by the 
intervening medium. 

The wide variations in pulse broadening seen within a single pulse are, however,
difficult to explain unless the scattering screen is illuminated differently 
by each component.  While we expect variation in the pulse
broadening from component to component and from pulse to pulse owing to their
being instantaneous samplings of the time variable diffractive effects,
why does the {\it minimum} exponential time scale of the components
at 1.4\,GHz agree with that extrapolated from 0.3\,GHz? 
We have considered the possibility that the
components of giant pulses are intrinsically single impulsive events 
that are multiply refracted and/or diffracted by discrete regions of enhanced
plasma density and/or turbulence, respectively. Lyne \& Thorne (1975) \nocite{lt75}
invoked similar intermittency in the wave front perturbing medium
to explain the irregular and rapidly varying
effects of the 1974 event, and Cordes, Hankins \& Moffett (1998\nocite{chm98}) have studied the effects of 
discrete, small scale ``screenlets'' on giant pulses. 
An intermittent scattering medium that multiply refracts and/or diffracts a
single impulsive signal appears to solve the problem, but has serious 
difficulties explaining the frequency independence of the spread and 
multiplicity of pulse components.

Consider a screen at a distance $xD$ from the pulsar and $(1-x)D$ from
the observer. Diffraction leads
to an expansion of the angular spectrum of the intensity by an angle
$\theta_{\rm d}\equiv\lambda/l_\circ$ where $l_\circ$ is the coherence
scale for one radian of phase difference across the wavefront. 
An impulse which passes through this screen is broadened by the
multipath diffractive time scale, $\tau_{\rm d}$. If the
scattering material covers a transverse scale of $l_{\rm d}=xD\theta_{\rm d}$ 
centered on the geometric line of sight, then 
$\tau_{\rm d}=x(1-x)D\theta_{\rm d}^2/2c$.  A measure of the apparent angular size 
of the pulsar is $\theta_{\circ,d}=x\theta_{\rm d}$. 
If $\tau_{\rm d}$ is 4\,$\mu$s at 1\,GHz and the screen is located 2 pc
from the pulsar, $x=10^{-3}$, $\theta_{\rm d}=60$\,mas, $\theta_{\circ,d}=60\,\mu$as,
$l_\circ=10^7$\,cm, and $l_{\rm d}=1.8\times 10^{11}$\,cm. If the diffracting material
covers only a fraction of $l_{\rm d}$, then the pulse broadening time will be
{\it reduced} relative to $\tau_{\rm d}$ owing to the reduction in the multipath
propagation (\cite{chm98}). 
The observed frequency scaling between 0.3-GHz and 
1.4-GHz broadening times reported above is not consistent with this result.  
In addition, a diffracting region
not located along the line of sight will result in components
disappearing from view as the frequency increases, owing to the reduced viewing
zone set by $l_{\rm d}$ centered on the line of sight. This is in conflict
with the current observational results.

Refraction in an intermittent medium has similar limitations to that
of diffraction. Consider a uniform density, spherical plasma lens at a transverse
distance $b$ from the line of sight with a transverse dimension significantly
less than $b$. The excess propagation delay from the pulsar to the observer
is dominated by the geometric delay $\tau_{\rm g}=b^2/(2cx(1-x)D,$ and is 
independent of frequency owing to the highly
aberrant lens. While multiple lenses of this form appear to satisfy the
frequency independence of giant pulse component spacings, they will {\it not}
satisfy the frequency independence of the number of components and their
overall distribution in longitude. 
More components over a wider range of longitude would be expected at lower frequencies
as is the case for diffraction discussed above. 

We conclude that the multiplicity of components is intrinsic to the pulsar
emission mechanism or to propagation effects within the pulsar magnetosphere. 
Multiple scatterings of a single emitted pulse component cannot easily explain
the observed spread and multiplicity of pulse components.  
Despite this, we favor propagation effects in the Crab Nebula as an explanation
for the shape of the components at 1.4\,GHz and below.
Alternatively, the observed component shape may be intrinsic to the 
emission.  Each component is emitted with its own time scale, but the 
characteristic $(t/a_3)\exp(-t/a_3)$ shape.  Components with time scales 
shorter than the interstellar broadening time scale are broadened by scattering
in the ISM. This model explains the observed correlation between the minimum 
time scale and the low-frequency scattering, but involves two separate
explanations for the characteristic shape of the components.
Determining whether the long-$\tau$ end of the distribution scales with the
low-frequency scattering or is independent of it would distinguish between the
two possibilities.  
Observations at multiple frequencies during times of minimal 
scattering at 0.3\,GHz are critical to further inquiry.

\subsection{Energies and Spectral Indices}

The distribution of pulse energy amplitudes from the 1.4/0.6-GHz experiment
are displayed along with the corresponding detection thresholds in Figure 
\ref{gbvla}. The 1.4-GHz amplitudes of the 13 pulses that were not detected 
at 0.6\,GHz are also shown.  
Solid lines corresponding to spectral indices $-2.2$ and $-4.9$ are shown, 
where spectral index $q$ is defined by $A_{\rm GB}/A_{\rm VLA} = (0.6/1.4)^q$.
The overall spectral index for the Crab pulsar is $-3.1$, while  
the spectral index for the average main pulse, which is shown as a dotted line 
in Figure \ref{gbvla} (\cite{m97}), is $-3.0$.
The pulse amplitude of the average main pulse
is 5.4 $\times$ 10$^{-3}$ Jy-s at 0.6\,GHz. The largest 0.6-GHz giant pulse 
therefore has a pulse amplitude of about 150 times the amplitude of the 
average pulse.  The giant pulses are narrower than the average pulse,
and so are even stronger relative to the average pulse within this window.


Lundgren \etal (1995)\nocite{lcu+95} concluded from their analysis of the 
relative contribution of giant pulses to the average as a function of 
frequency ({\it cf.} section \ref{secintro}) 
that if the emission is narrow band, the rate of giant
pulses must increase with frequency, while if the emission is broadband, 
then the giant pulses must have flatter spectra than the weak pulses.
Heiles \& Rankin (1971)\nocite{hr71} found that their measured 
spectral indices at low radio frequencies ranged from nearly 0 to less than 
$-3.0$.
Similarly, Moffett (1997)\nocite{m97} found that between 1.4 and 4.9\,GHz, 
giant pulse spectral indices ranged from 0 to $-4$, with an average of about 
$-2$. These two studies corroborate Lundgren's analysis.
We find that at least 70\% of the pulses are broadband, and so we expect that 
their spectral indices are, on average, flatter than the average main pulse 
spectral index. 

The average spectral index of the giant pulses in Figure 7 which were detected
at both frequencies is $-3.4$, which
is comparable to that of the average main pulse.  This 
estimate is biased by the fact that the pulses that were not detected
at 0.6 GHz have flatter spectral indices. 
In addition there may be systematic errors in
flux calibration that could change the average spectral index by up to
$0.4$.   

The individual giant pulse spectral indices display a relatively large 
scatter.  Individual pulse spectral indices are known for two other pulsars.  
The distributions of spectral indices for pulsars B0329$+$54 and B1133$+$16 
have a standard deviation of 0.2, and range from -1.6 to -3.1 and -2 to 0,
respectively (\cite{bs78}).
One contribution to the spectral index variations is the stochastic uncertainty 
in the determination of the amplitudes that is introduced by the low number
of degrees of freedom in narrow band observations of intrinsically short
duration pulses.  We estimate this {uncertainty} to be of order 10\%  (20\%)
at 1.4 (0.6) GHz for an intrinsic pulse width of 1\,$\mu$s. 
This is not large enough to explain the scatter in the spectral indices.
The scatter could also be intrinsic to the radiation emission process. 
The signal could consist of a randomly occurring series of nanosecond 
impulses whose combined power spectrum is irregular.  This would
cause scatter in the observed spectral indices.
But with the radiation extending over 1\,$\mu$s or more, there are many 
nanosecond pulses which would smooth out this distribution. 
Alternatively, one might expect the spectral index to vary 
due to properties of the emission beam: frequency dependence and orientation
with respect to the observer.  Future observations with higher time
resolution and also with many samples across
the spectrum will provide critical new insight into the giant pulse
emission process.

\subsection{Models of the Emission Beam}
\label{secwithwidth}

We consider two  general models for giant pulse emission.  In the
temporal model the giant pulse emission components are
impulses in time ($<$1\,$\mu$s) with angular beam widths comparable to 
that of the average pulse, $\sim 3^\circ$.
Following Lundgren (1994)\nocite{lun94}, we also consider a model in which the 
enhanced emission during a giant pulse results from a steady narrow beam
($<$1$^\prime$) whose position wanders on time scales 300\,$\mu$s 
$\ll t \ll P$.  In this case intrinsic
structure within a particular giant pulse is due to
structure within this narrow beam.
In both models, the average beam may be 
circular, as in polar cap models, or fan-shaped, as is likely if the 
emission comes from the outer gaps.  

In the angular model, the narrow beam of emission may wobble in either the
$l$ direction (along the trajectory of the line of sight),
or the $\phi$ direction (perpendicular to the trajectory
of the line of sight).
The width of the giant pulses corresponds to the size of the 
beam in this model, while the jitter in arrival times $\sigma_{\rm toa}$
corresponds to the wobbling of the beam along $l$.
For giant pulses which occur a fraction $f$ of the
time, the wobble in $\phi$ is then $w_{\rm gp}/(P f)$
where $P$ is the pulse period (following Lundgren \etal 1995)\nocite{lcu+95}.
Lundgren was able to separate the giant pulse and normal pulse distributions
at 0.8\,GHz, and found that one of 40 pulses is giant. 
Since the giant pulses form a separate distribution, then if they are all 
broadband, they will all also appear at higher radio frequencies.  
Then at 1.4\,GHz or
4.9\,GHz, one of 40 pulses should be giant.  In fact, Moffett (1997) finds that 
one of 50 pulses at 1.4\,GHz has an energy greater than 20 times the average. 
At 4.9\,GHz the intrinsic width of
the giant pulses is 0.1\,$\mu$s to 0.4\,$\mu$s.
We find $\sigma_{\rm toa} \approx 100\,\mu$s, so  
the $0.001^\circ$ to $0.004^\circ$ beam must wobble $1.1^\circ$ in $l$
and $0.05^\circ$ to $0.2^\circ$ in $\phi$.  This is not consistent with a 
narrow beam wobbling within a roughly circular average beam.  

\subsection{The Emission Mechanism}
\label{secemission}

Radio emission from pulsars must come from a coherent emission process 
(\cite{cor81}). The exact process is very uncertain, as is the location(s) of 
the emission. It is not necessary for the giant pulse emission to originate
at the same place or in the same way as the ordinary pulse emission.  The
broadband nature of the giant pulse emission provides the main constraint on
its origin.  
According to Melrose (1996)\nocite{mel96}, broadband emission is 
traditionally associated with models in which the emission occurs
at a pair production front in the polar cap, or by Schott radiation
from a corotating charge and current distribution outside the light cylinder
({\em e.g.}, \cite{dk85}; Ardavan 1992\nocite{a92}, 1994\nocite{a94}).  
The emission process itself could rely on plasma instabilities (\cite{cr77};
\cite{a93}; \cite{mu79}; \cite{k+91}; \cite{wea96}).
Alternatively, other maser processes such as linear acceleration emission 
(\cite{mel78}; \cite{row95}) or maser curvature emission (\cite{lm92}, 
1995\nocite{lm95}) could produce the radiation.
In any case, if the giant pulses are a temporal effect, this
variability in radio emission could be the result of
the statistics of a small number of coherently emitting regions which are
incoherently summed, or an increase in the coherence within a single emission 
region. 

Although we believe that the 
$(t/a_3)\exp(-t/a_3)$ characteristic shape of the giant pulse components at 
1.4\,GHz is most likely the effect of propagation through
the Crab nebula interface, the remaining questions ({\it cf.} section
\ref{secinterp}) lead to consideration of effects intrinsic to the pulsar.
An asymmetric shape is not expected for a simple narrow beam with an angular
wobble.  In either the temporal or angular beam model, this shape might be
explained by effects that occur as the signals traverse the pulsar magnetosphere
(Eilek 1998\nocite{eil98}). The effects of aberration
are too small to produce the broadening and asymmetry seen in these pulses,
if one confines the range of emission altitudes to 4.5 km, the limit 
obtained from the timing residual differences at the two frequencies.
In the temporal model, the asymmetric shape is consistent with any 
emission process which turns on with a rapid nearly linear rise, then
saturates and decays.  In this case one might expect the peak energy
to be independent of width, whereas we have seen that it is the total
pulse energy which is independent of width.  

If the model is truly temporal, then the angular size of the beam does not
affect the observed pulse width.  The radiation is beamed into a beam width
$\theta \approx  \gamma^{-1}$. The beam must be wider 
than any given 1.4-GHz pulse component, 50\,$\mu$s, and therefore 
$\gamma${\lower.5ex\hbox{$\; \buildrel < \over \sim \;$}}100. 
We can use this value of $\gamma$ to
estimate particle properties using a simplistic model of coherent curvature radiation. 
The power lost by the $N$ excess charged particles in the bunch will be 
$$
P_{\rm curv} = N^2 \left(\frac{2 e^2 \gamma^4 c}{3 \rho_{\rm c}^2}\right),
$$
where $e$ is the charge on an electron, $\gamma$ is the relativistic factor
$(1-v^2/c^2)^{-1/2}$, and $\rho_{\rm c}$ is the radius of curvature of the 
magnetic field.  We observe $6.3 \times 10^{-3}$ Jy-s in a 50-MHz band, so 
$P_{\rm curv}$ must equal the measured luminosity, which is therefore greater 
than $3.9 \times 10^{23}$ erg s$^{-1}$, assuming
a distance of 2 kpc, and a circular beam 300\,$\mu$s = 3$^\circ$  wide.  Then
the number of particles in the bunch must be at least 
$$
N=9.2 \times 10^{19} \left(\frac{\gamma}{100}\right)^{-2}\left(
\frac{\rho_{\rm c}}{10^8 \rm ~cm}\right).
$$
These particles must fit within a cube with volume 
$\le \lambda_{\rm obs}^3$ in order to maintain the observed coherence, so the 
number density of excess charges must be 
at least
$\delta n_{\rm e} = N/\lambda_{\rm obs}^3 = 
9.9 \times 10^{15} ({\gamma \over 100})^{-2}$\,cm$^{-3}$ 
for the parameters used above, and a wavelength of 21\,cm.  For comparison,
the Goldreich-Julian density in the observer's frame is
${n_{\rm G-J}} ={\Omega B / e 2 \pi c } = 8.3 \times 10^{12} 
({R \over R_{\rm NS}})^{-3}$\,cm$^{-3}$, 
using a surface magnetic field of $4 \times 10^{12}$\,G.
The excess charge density can be
further reduced if several bunches are radiating in a periodic structure.  

\section{Summary}

The giant pulse emission from the Crab pulsar is broadband since
70\% of the pulses are observed in our 1.4/0.6-GHz experiment. The strong 
correlation in arrival times at the two frequencies
implies that the same radiating unit is operating at both frequencies.  
The giant pulses
display a scatter in spectral index that is consistent with or flatter than the 
spectral index of the average main pulse component.  Pulsar emission
models are restricted to those that can explain the broadband nature
of the giant pulse radiation on intrinsic observed time scales of one
to 10 microseconds.

Above 1\,GHz a multiplicity of components
is observed with frequency independent spacing and number. The 1.4/0.6-GHz data
are also consistent with frequency independence. We conclude that the
multiplicity is intrinsic to the pulsar emission process and not the
result of multiple imaging in the intervening plasma.

The exponential pulse broadening time scale of the average pulse at 0.3\,GHz
and of the giant pulses at 0.6\,GHz are consistent with multipath propagation
effects. The large values and rapid variations indicate a special scattering
region. We identify this region with the interface between the Crab synchrotron
nebula and the surrounding supernova ejecta.
At 1.4\,GHz the observed giant pulse component shapes are characterized by 
fast rise and exponential decay time scales that are correlated. The 
minimum time scale is consistent with extrapolation of the pulse broadening
at lower frequencies with a filled turbulent scattering screen. 
The shape and the distribution of broadening at 1.4-GHz
are not fully understood. Multi-frequency simultaneous observations of 
giant pulses with higher time resolution
at epochs of minimal scattering at low frequency will provide
critical new insights into the emission processes and subsequent propagation
effects.

\acknowledgments

We thank John Ford for assistance with the trigger reception software.
Thanks to the observatory staff at both sites for minimizing internet 
traffic during the two hours of the experiment.
T.H. Hankins and D.A. Moffett thank the NSF for partial funding of 
this work under grant AST9315285. D.C. Backer and S. Sallmen thank
the NSF for partial funding of this work under grant AST9307913.
We are indebted to NASA for maintenance of the 25-m telescope at Green Bank.
We thank
J. Cordes, R. Jenet, M. Kramer, A. Melatos, and J. Weatherall for 
conversations and comments on the early versions of the paper, and the
referee for attentive reading and useful criticism.

\clearpage

\figcaption[Single Crab Giant Pulse at 1.4\,GHz]{
The top panel displays the total intensity $I$, along with linear and
circular polarizations $L$ and $V$ for a single giant pulse at 1.4\,GHz, taken
at the VLA on 1996 May 21, with a temporal resolution of 0.5\,$\mu$s.
The vertical scale indicates
that this pulse reached a peak flux of 3000 Janskys.  The lower panel
indicates the linear polarization position angle wherever $L > 3$ times the off pulse rms noise.  
\label{vlabest}}

\figcaption[Single Crab Giant Pulse at 0.6\,GHz]{
The top panel displays the relative values at 0.6\,GHz of the total 
intensity $I$, and the linear and circular polarizations $L$ and $V$ 
for the same single pulse as shown in Figure \ref{vlabest}, taken with 
the GBPP.  The temporal resolution is approximately $34\,\mu$s.
The peak flux for this pulse was $\approx$7000\,Jy.
The negative and positive features on either side of the main peak are
artifacts due to the non-linear response of the GBPP.  The lower panel
indicates the linear polarization position angle wherever $L > 3$ times the off pulse rms noise.  
\label{gbbest}}

\figcaption[Correlation in Arrival Times of 1.4-GHz and 0.6-GHz Giant Pulses]{
The top panel displays the 1.4-GHz timing residuals against the 
0.6-GHz timing residuals.  The solid line passes through the origin with 
slope 1.  The bottom panel displays the same data with this line removed.
\label{resids}}

\figcaption[Component Fitting to a 1.4-GHz Giant Pulse]
{
An example of a simple 1.4-GHz single pulse profile.  
The intensity data are modelled by the dark solid line, which is created 
using the fitted components represented by the dashed lines.  
These components are characterized by a fast, nearly linear rise, 
followed by an exponential decay. The narrow component of the giant 
pulse shown here has a characteristic time scale of 1.2\,$\mu$s.
The fit residuals are shown in the lower panel with the $2\sigma$ 
uncertainty envelope shown, where $\sigma = (T_{\rm sys} + T_{\rm pulsar}) (\Delta \nu\,\Delta \tau)^{-1/2}$, and $\Delta \nu$, $\Delta \tau$ are the receiver bandwidth and post-detection integration time, respectively.
\label{pulse69}}

\figcaption[Component Fitting to a 1.4-GHz Giant Pulse]{
An example of a complex 1.4-GHz single pulse profile.  
The intensity data are modelled as described for Figure \ref{pulse69}.
The final component of the giant pulse shown here has a 
characteristic time scale of 5.7\,$\mu$s. The fit residuals are as described for Figure \ref{pulse69}.
\label{pulse56}}

\figcaption[Simultaneous Giant Pulses at 5\,GHz and 1.4\,GHz]{
(a) A single giant pulse recorded at
4.8851\,GHz, plotted with 1-$\mu$s resolution.  
(b) The same pulse recorded at 1.4351\,GHz, plotted with the same time 
resolution.
\label{5GHz}}

\figcaption[Energy Correlation of Dual Frequency Giant Pulses]{
Pulse energy amplitudes in Jy-s at 0.6\,GHz and 
1.4\,GHz.  The solid
circles denote the 29 pulses which were detected at both frequencies.  
Error bars reflect the measurement {uncertainty}, which is negligible for the
1.4-GHz data.  Uncertainty in the telescope gain 
calibration used at 0.6\,GHz 
introduces an additional systematic uncertainty of 50\%.  The
open circles represent those pulses seen at 1.4\,GHz which were certainly
not detected at 0.6\,GHz.  The horizontal dashed line represents the
cutoff of 0.075 Jy-s, below which we could not detect pulses at 0.6\,GHz.  The
vertical dashed line indicates our estimate of the VLA threshold corresponding
to 6 times the rms noise.  The one pulse with a 1.4-GHz energy less than 
this occurred while our threshold was set to 5 times the rms noise.  The 
solid lines represent spectral indices $q=-2.2$ and $q=-4.9$.  The dotted
line indicates the average main pulse spectral index.
\label{gbvla}}

\clearpage
\epsscale{0.6}
\plotone{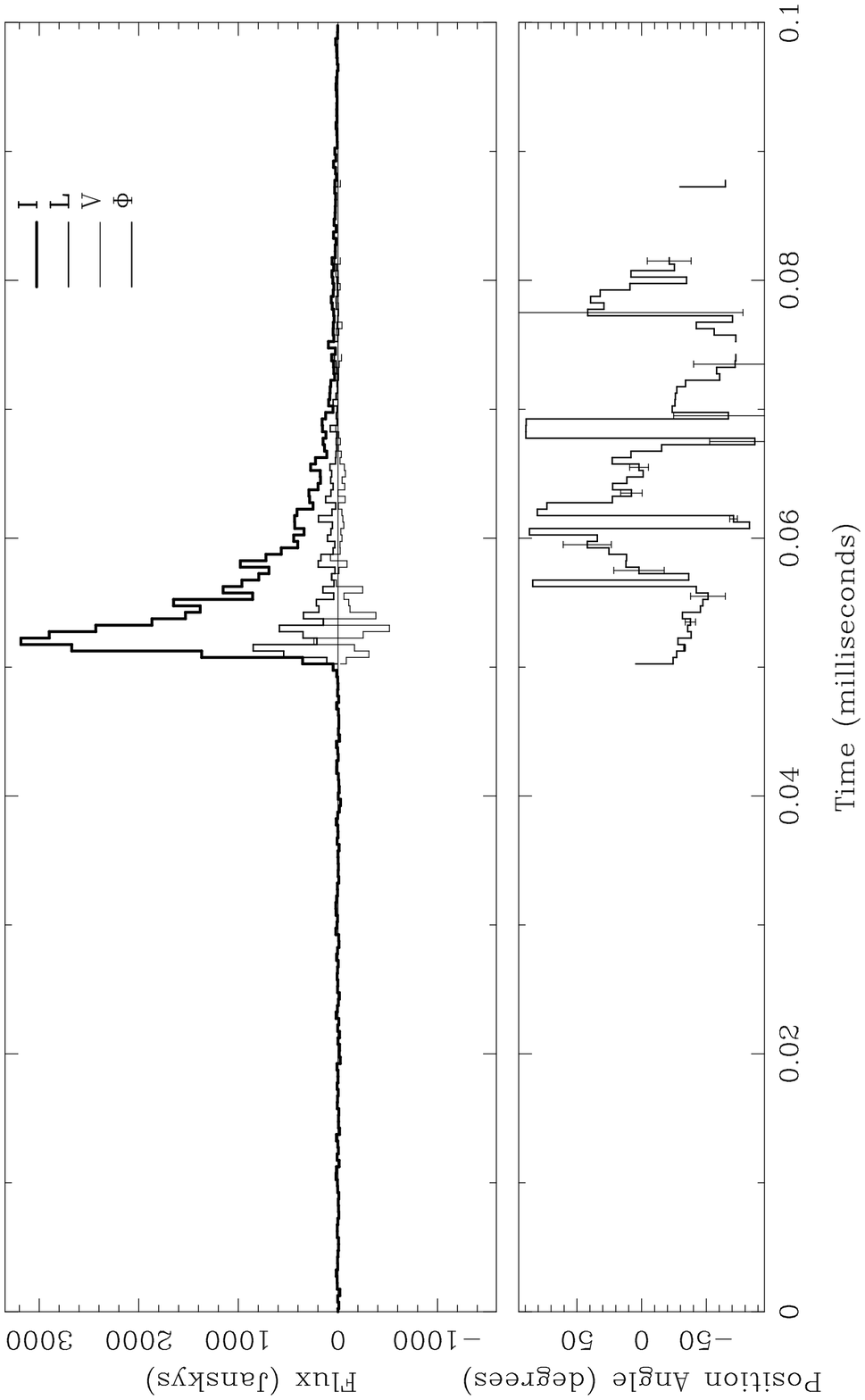}
\clearpage
\epsscale{0.8}
\plotone{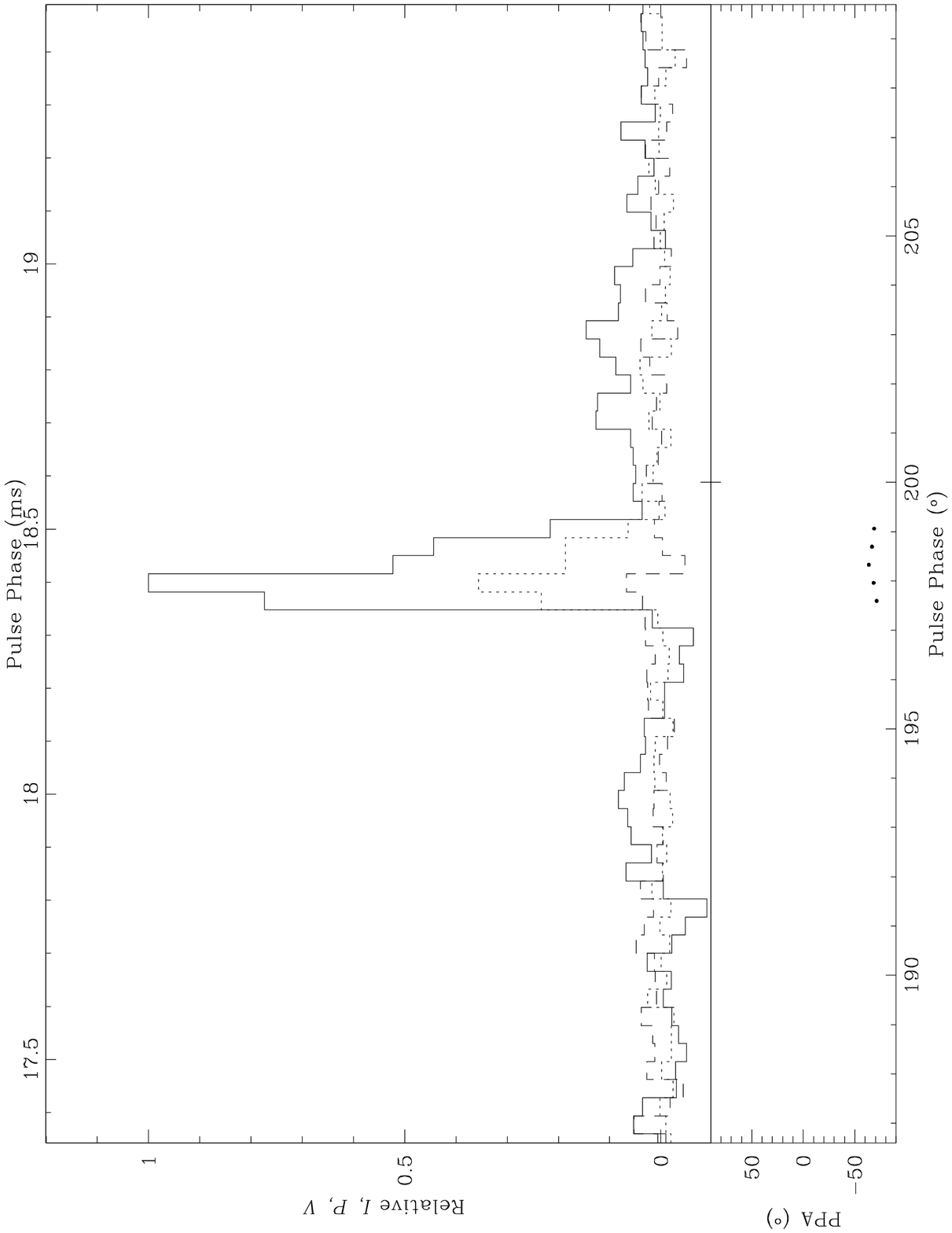}
\epsscale{1.0}
\clearpage
\plotone{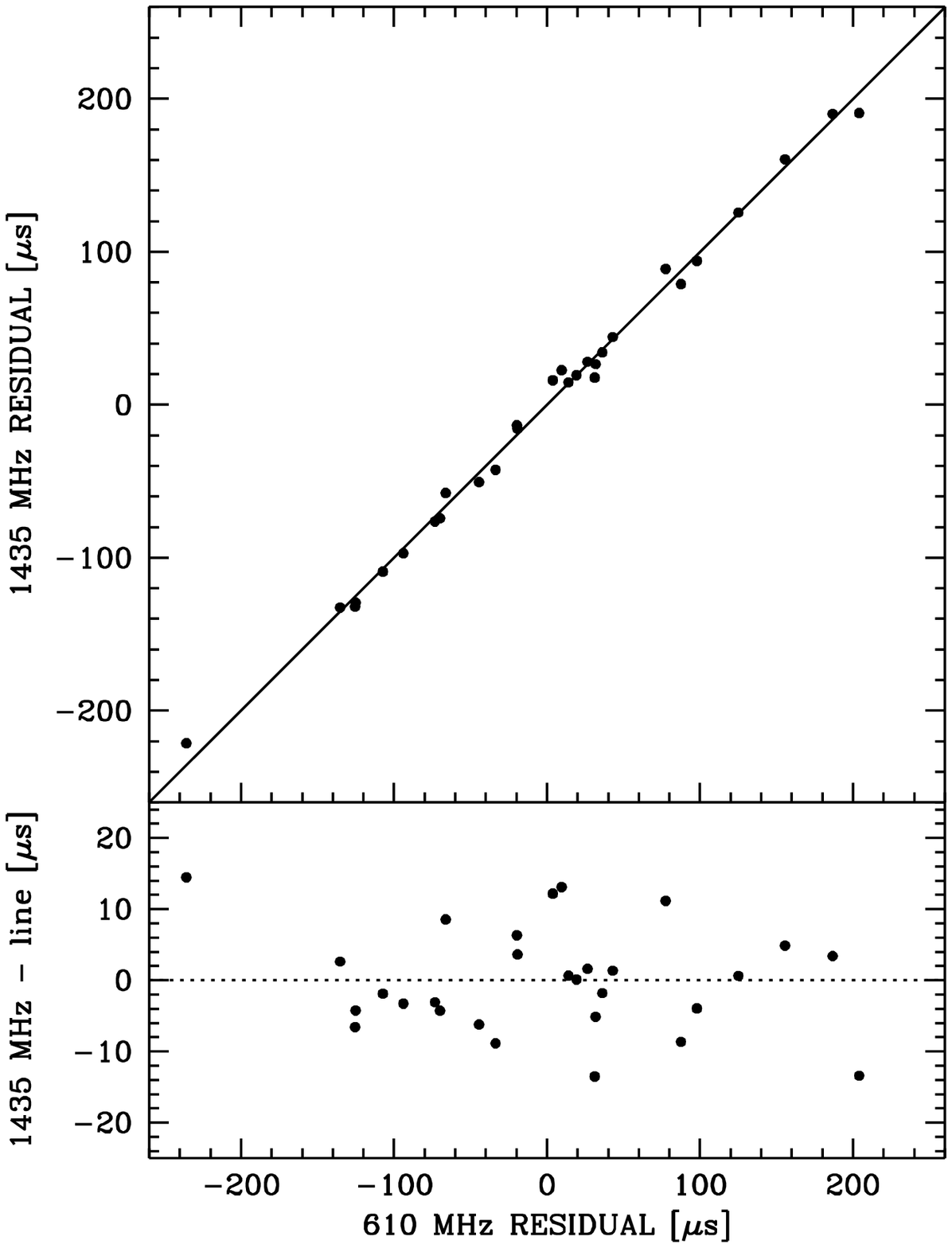}
\clearpage
\plotone{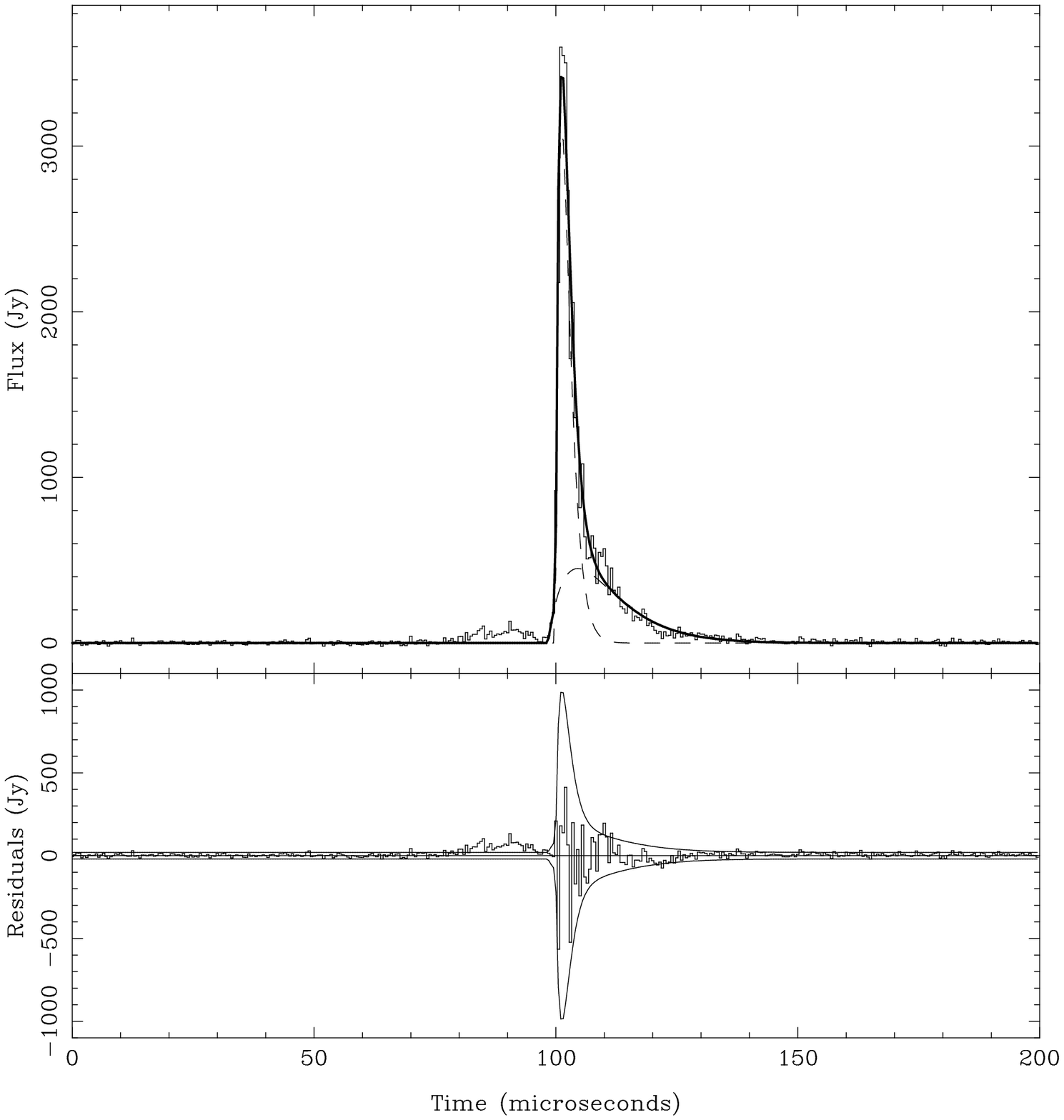}
\clearpage
\plotone{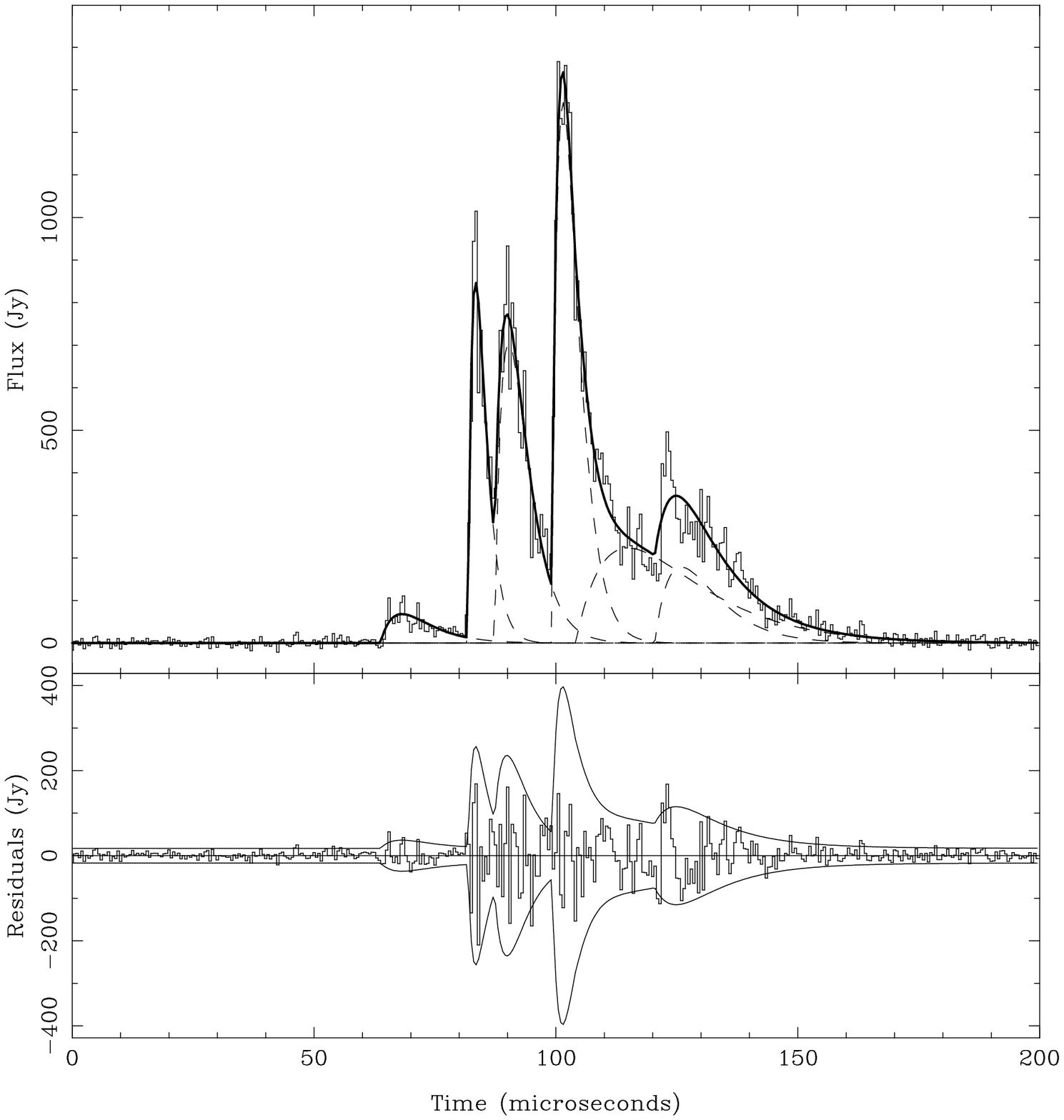}
\clearpage
\epsscale{0.4}
\plotone{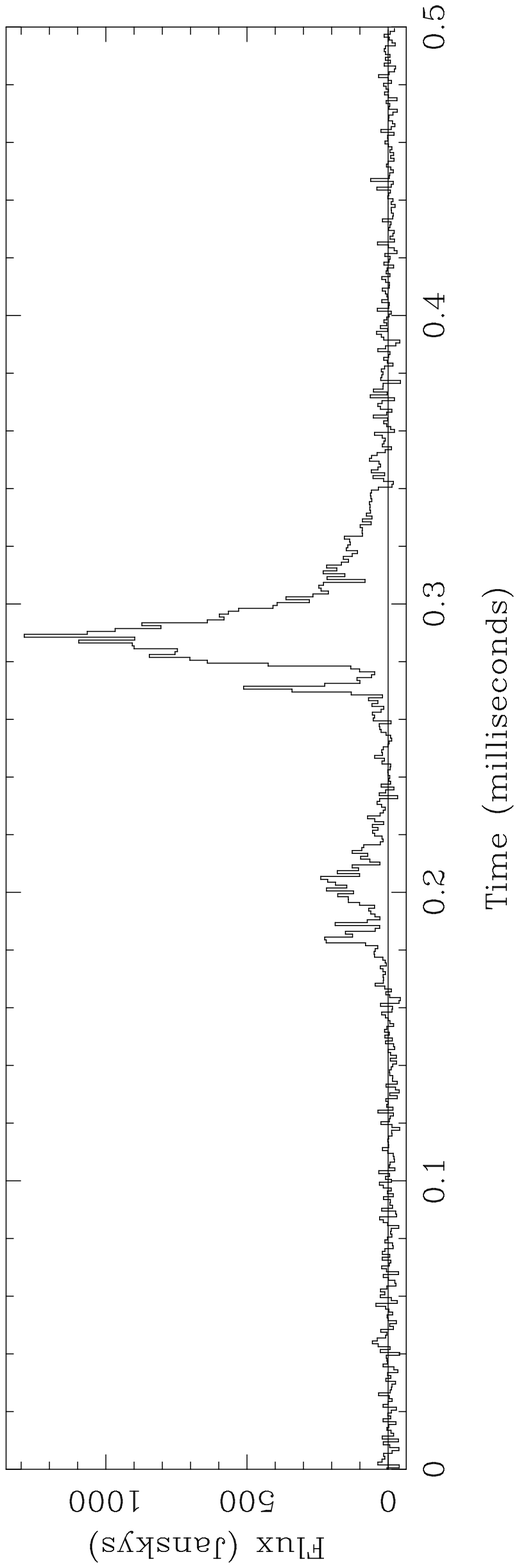}
\epsscale{1.0}
\clearpage
\epsscale{0.4}
\plotone{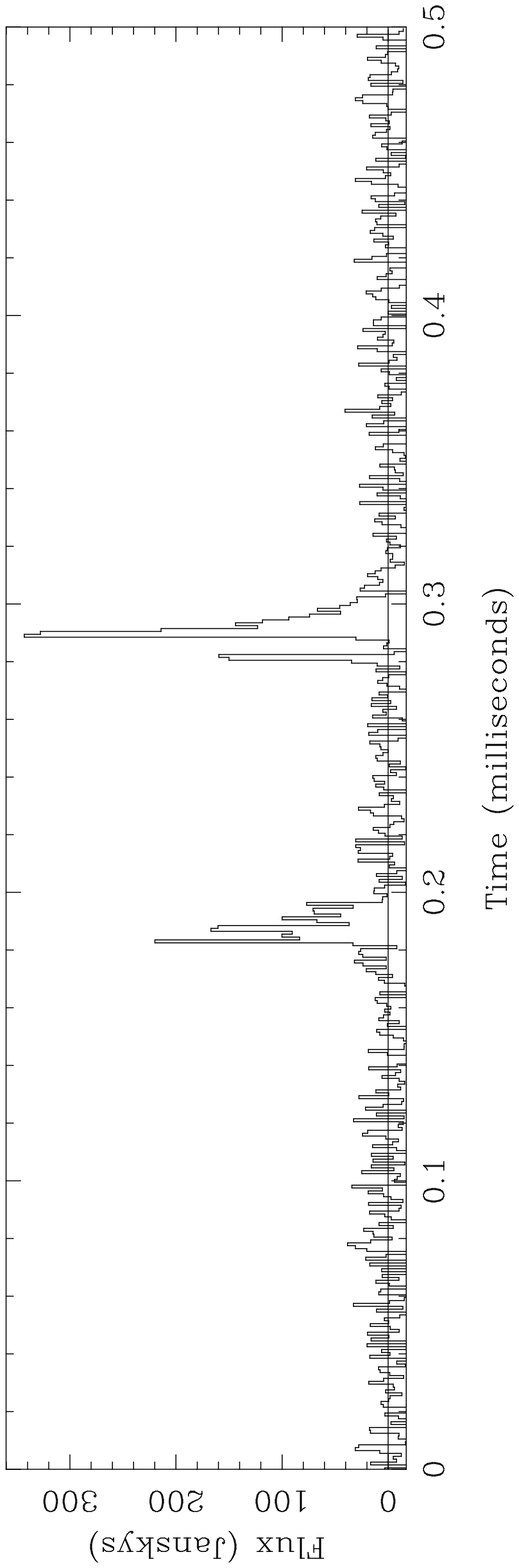}
\epsscale{1.0}
\clearpage
\plotone{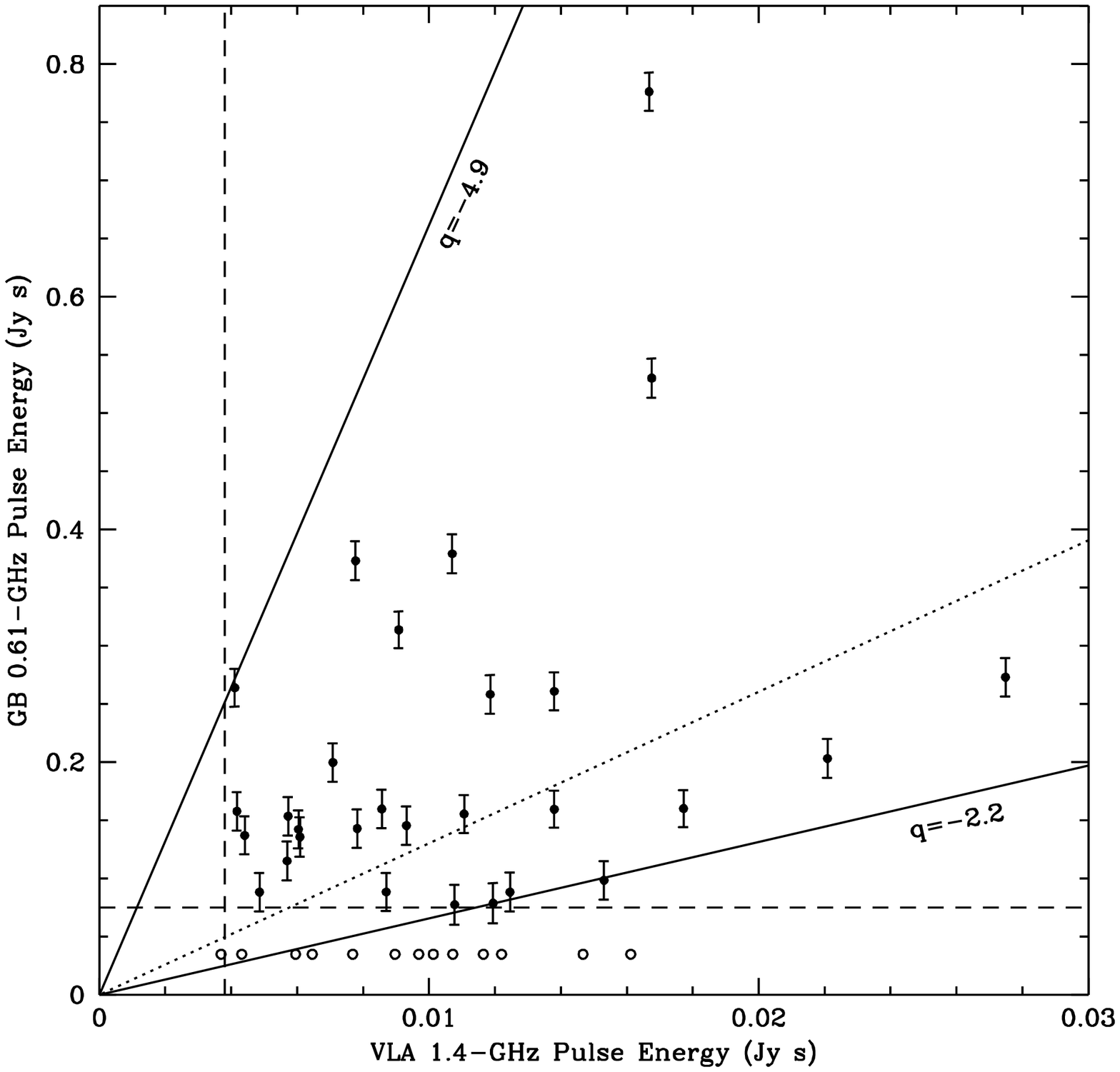}

\end{document}